\documentclass[useAMS,usenatbib]{mn2e}
\usepackage[T1]{fontenc}
\usepackage{aecompl}

\usepackage{amsmath,graphicx,amsfonts,amssymb}
\usepackage{natbib}

\newcommand{\figref}[1]{figure~\ref{#1}}

\renewcommand{\eqref}[1]{equation~(\ref{#1})}
\newcommand{\eqsref}[1]{equations~(\ref{#1})}

\newcommand{\exref}[1]{(\ref{#1})}

\newcommand{\bea}{\begin{eqnarray}}
\newcommand{\eea}{\end{eqnarray}}
\newcommand{\beq}{\begin{equation}}
\newcommand{\eeq}{\end{equation}}
\newcommand{\lt}{\left}
\newcommand{\rt}{\right}

\renewcommand{\la}{\langle}
\newcommand{\ra}{\rangle}

\newcommand{\tA}{\tau_{\rm A}}
\newcommand{\tAg}{\overline{\tau}_{\rm A}}
\newcommand{\tnl}{\tau_{\rm nl}}
\newcommand{\tnlg}{\overline{\tau}_{\rm nl}}

\newcommand{\vskipfig}{}
\newcommand{\dd}{\partial}
\newcommand{\vB}{\mathbf{B}}
\newcommand{\vBloc}{\vB_{\rm loc}}
\newcommand{\vu}{\mathbf{u}}
\newcommand{\vb}{\mathbf{b}}
\newcommand{\vbloc}{\hat{\vb}_{\rm loc}}
\newcommand{\vz}{\mathbf{z}}
\newcommand{\vr}{\mathbf{r}}
\newcommand{\vrp}{\vr_\perp}

\newcommand{\vup}{\vu_\perp}
\newcommand{\vbp}{\vb_\perp}
\newcommand{\vzp}{\vz_\perp}

\newcommand{\dzp}{\delta z_\perp}
\newcommand{\dzg}{\overline{\delta z}_\perp}
\newcommand{\tdzp}{\delta\tilde z_\perp}
\newcommand{\dvzp}{\delta \vz_\perp}

\newcommand{\dbp}{\delta b_\perp}

\newcommand{\lpar}{l_\parallel}

\newcommand{\du}{\delta u}

\newcommand{\kperp}{k_\perp}
\newcommand{\kpar}{k_\parallel}

\title[Refined critical balance]{Refined critical balance in strong Alfv\'enic turbulence}
\author[A.~Mallet, A.~A.~Schekochihin, and B.~D.~G.~Chandran]{
A.~Mallet,$\!^{1,2,3}$
A.~A.~Schekochihin,$\!^{1,4}$
and B.~D.~G.~Chandran,$\!^{3,4}$\\
$^{1}$Rudolf Peierls Centre for Theoretical Physics, University of Oxford, 1 Keble Rd, Oxford OX1 3NP, UK\\
$^{2}$Wolfgang Pauli Institute, Faculty of Mathematics, University of Vienna,
Oskar-Morgenstern-Platz 1, 1090 Vienna, Austria\\
$^{3}$Space Science Center and Department of Physics, University of New Hampshire, Durham, NH 03824, USA\\
$^{4}$Merton College, Oxford OX1 4JD, UK}

\begin{document}

\date{\today}

\pagerange{\pageref{firstpage}--\pageref{lastpage}} \pubyear{2014}

\maketitle

\label{firstpage}

\begin{abstract}
We present numerical evidence that in strong Alfv\'enic turbulence, 
the critical balance principle---equality of the nonlinear 
decorrelation and linear propagation times---is scale invariant, in the 
sense that the probability distribution of the ratio of these times is independent 
of scale. This result only holds if the local 
alignment of the Elsasser fields is taken into account in calculating the nonlinear 
time. At any given scale, the degree of alignment is found to increase with 
fluctuation amplitude, 
supporting the idea that the cause of alignment is mutual dynamical 
shearing of Elsasser fields. 
The scale-invariance of critical balance 
(while all other quantities of interest are strongly intermittent, i.e., 
have scale-dependent distributions)
suggests that it is the most robust of the 
scaling principles used to describe Alfv\'enic turbulence.   
The quality afforded by situ fluctuation measurements in the solar wind  
allows for direct verification of this fundamental principle.  
\end{abstract}


\section{Introduction}

Strong plasma turbulence is present in many astrophysical systems, 
and is directly measured by spacecraft in the solar wind \citep{Bruno05}. 
The precision and sophistication achieved by these measurements in the 
recent years have enabled direct observational testing of theories of magnetized 
plasma turbulence that go beyond crude dimensional scalings---we mean, in particular, 
measurements of spatial anisotropy \citep{Horbury08,Podesta09,Wicks10,Chen11}, 
intermittency \citep{Horbury97,Marsch97,Carbone04,Salem09,Zhdankin12,Osman14} 
and alignment \citep{Podesta09jgr,Chen12,Wicks13prl,Wicks13apj} 
of magnetic and velocity fluctuations. In this Letter, we report a new result, obtained 
numerically, that elicits a striking but physically plausible relationship between 
these three aspects of the structure of plasma turbulence.   

In a strong mean magnetic field $\mathbf{B}_0$, Alfv\'enic 
fluctuations decouple from compressive ones and satisfy the reduced magnetohydrodynamic 
(RMHD) equations, which correctly describe Alfv\'enic turbulence 
in both strongly and weakly collisional plasmas \citep[see, e.g.,][and references therein]{Sch09}. 
The equations are best written in \citet{Elsasser50} variables 
$\vzp^\pm = \vup \pm \vbp$, where $\vup$ and $\vbp$ are the velocity and 
magnetic-field  (in velocity units) perturbations, perpendicular to $\mathbf{B}_0$:
\beq
\dd_t \vzp^\pm \mp v_A \dd_z \vzp^\pm + \vzp^\mp \cdot \nabla_\perp \vzp^\pm = -\nabla_\perp p,
\label{eq:RMHD}
\eeq
where the pressure $p$ is determined via $\nabla_\perp \cdot \vzp^\pm = 0$, 
$v_A=|\vB_0|$ is the Afv\'en speed, and $\vB_0$ is in the $z$ direction. 

The modern understanding of the small-scale structure of Alfv\'enic turbulence described by 
\eqsref{eq:RMHD} (and, indeed, the validity of these equations) rests on the 
fluctuations being spatially anisotropic with respect to 
the magnetic field, and ever more so at smaller scales---this is supported both by solar-wind measurements 
and by numerical simulations (see, e.g., \citealt{Chen11} and references therein). 
The relationship between the parallel and perpendicular coherence scales 
of the fluctuations is set via the {\em critical balance} conjecture \citep{GS95}, 
whereby the nonlinear-interaction and the Alfv\'en-propagation times,
\beq
\tnl^\pm \doteq \frac{\lambda}{\dzp^\mp \sin{\theta}}, 
\quad
\tA^\pm\doteq \frac{\lpar^\pm}{v_A},
\label{eq:tnl}
\eeq
are expected to be comparable at each scale in some, shortly to be discussed, statistical sense. 
The Alfv\'en time is related 
solely to the scale $\lpar^\pm$ of the the fluctuations along the magnetic field, 
while the nonlinear time depends on the fluctuation amplitudes $\dzp^\pm$, 
their scale $\lambda$ perpendicular to the field and on the angle $\theta$ between 
$\dvzp^+$ and $\dvzp^-$---when this angle is small, the nonlinearity 
in \eqsref{eq:RMHD} is weakened, which is why we have included $\sin\theta$
in the definition of $\tnl^\pm$. This effect that can become increasingly 
important at smaller scales as envisioned by the ``dynamic alignment'' 
conjecture \citep{Boldyrev06,Mason06} (its small-scale validity is, however, 
disputed in \citealt{Beresnyak11,Beresnyak12}). 

Both the dynamics of weak turbulence ($\tA^\pm\ll\tnl^\pm$) and the causal 
impossibility to maintain $\tA^\pm\gg\tnl^\pm$ (fluctuations 
in planes perpendicular to $\vB_0$ separated by a distance $l$ decorrelate if 
$l$ greatly exceeds the distance an Alfv\'en wave can travel during one nonlinear time, 
$l \gg v_A\tnl^\pm$) push the two time scales 
towards critical balance \citep{GS97,Boldyrev05,Nazarenko11}. 
This guarantees strong turbulence, 
with cascade time $\tau_c^\pm \sim \tnl^\pm \sim \tA^\pm$. Then, 
by the Kolmogorov argument, the scale independence of the energy fluxes,
\beq
\varepsilon^\pm \sim \frac{(\dzp^\pm)^2}{\tau_c} 
\sim \frac{(\dzp^\pm)^2 v_A}{\lpar^\pm}
\sim \frac{(\dzp^\pm)^2\dzp^\mp\sin\theta}{\lambda}
\sim \text{const},
\label{eq:eps}
\eeq
immediately implies $\dzp^\pm \propto (\lpar^\pm)^{1/2}$, 
or, equivalently, the ``parallel energy spectrum'' $E(\kpar) \propto \kpar^{-2}$, 
indeed seen in both the solar wind and simulations 
\citep{Horbury08,Podesta09,Wicks10,Chen11,Beresnyak14par}. 
The perpendicular scaling $\dzp^\pm \propto \lambda^\alpha$ 
is harder to establish as it depends on the scaling of $\sin\theta$---there 
is a continued debate whether the numerical evidence 
supports dynamic alignment ($\alpha=1/4$, \citealt{Perez12}) 
or (at small enough scales) does not ($\alpha=1/3$, \citealt{Beresnyak14}).  

As the resolution of such debates depends 
crucially on measuring precise scaling exponents, it is important 
to put the scaling formalism outlined above on a more precise 
footing. Indeed, what does ``$\sim$'' precisely mean in 
relations such as \eqref{eq:eps}? And how does one derive precise scaling 
laws on the basis of such relations?---precise in the sense 
of definite predictions about unambiguously defined statistical 
averages calculated from an ensemble (or a time history) of 
random solutions of \eqref{eq:RMHD}. 

That this is not a trivial question has long been known in the 
older field of hydrodynamic turbulence, where the 
statement $\varepsilon\sim \du^3/\lambda \sim\text{const}$, analogous 
to \eqref{eq:eps} ($\du$ are velocity increments), 
does not imply $\la\du^n\ra\propto\lambda^{n/3}$ 
for any moment except $n=3$---a phenomenon of {\em intermittency} 
of turbulent fluctuations \citep{Frisch95}. Furthermore, 
$\varepsilon$ is also an intermittent quantity: apart from $\la\varepsilon\ra$,  
no other moment of $\varepsilon$ is scale-independent.
Then ``$\varepsilon \sim \du^3/\lambda$'' means 
that both sides have the same distribution,  
which depends on $\lambda$ (``refined similarity hypothesis,'' \citealt{Kolmogorov62}). 
We will adopt the same approach to \eqref{eq:eps}, noting that, 
in Alfv\'enic turbulence, not only the amplitudes $\dzp^\pm$, but also $\lpar^\pm$ 
and $\theta$ (all precisely defined below) are intermittent (have 
distributions that depend on $\lambda$ {\em in a non-self-similar way})   
and mutually {\em dependent} random variables. 

In what follows, we will examine the joint statistical distribution 
of $\dzp^\pm$, $\lpar^\pm$ and $\theta$ as a function of $\lambda$ 
and show that critical balance is 
a more robust statistical statement than any other of the ``$\sim$'' 
relations---in the sense that the nonlinearity parameter 
\beq
\chi^\pm \doteq \frac{\tA^\pm}{\tnl^\pm} = 
\frac{\lpar^\pm\dzp^\mp\sin{\theta}}{v_A\lambda},
\label{eq:chi}
\eeq
while still a random variable, has a distribution that is 
independent of scale. We call this statement, 
which in the ``$\sim$'' language could be written as $\chi^\pm\sim1$, 
the {\em refined critical balance (RCB)}. 
We interpret it as evidence that critical balance results from a dynamical process 
that happens to inertial-range fluctuations in a completely scale-invariant way. 
The presence of the alignment angle $\theta$ 
in \eqref{eq:chi} will turn out to be an essential feature of the RCB. 
We will also examine how the (non-scale-invariant) distributions 
of $\tA^\pm$ and $\tnl^\pm$ combine to give rise to a scale-invariant $\chi^\pm$. 

\section{Definitions} 

We first define the quantities 
of interest. The fluctuation amplitudes are measured by 
increments 
\beq
\dzp^\pm \doteq |\dvzp^\pm| \doteq |\vzp^\pm(\vr_0+\vrp) - \vzp^\pm(\vr_0)|, 
\quad
\lambda \doteq |\vrp|,
\label{eq:dz}
\eeq
where $\vr_0$ is an arbitrary point (irrelevant under averaging 
because turbulence is homogeneous) and $\vrp$ the separation in 
the plane perpendicular to $\vB_0$ (moments of $\dzp^\pm$ only 
depend on $\lambda$ because of global isotropy in the perpendicular plane). 
The alignment angle is given by 
\beq
\sin\theta \doteq \frac{|\dvzp^+\times\dvzp^-|}{\dzp^+\dzp^-}. 
\label{eq:theta}
\eeq
The parallel coherence length $\lpar^\pm$ corresponding to 
a perpendicular separation $\vrp$ is defined 
as the shortest distance along the {\em perturbed} field line at which the Elsasser-field 
increment is the same as $\dzp^\pm$ \citep{Cho00,Maron01,Matthaeus12}:  
\begin{align}
\nonumber
&\lt|\vzp^\pm\lt(\vr_0 + \frac{\vrp + \lpar^\pm\vbloc}{2}\rt) - 
\vzp^\pm\lt(\vr_0 + \frac{\vrp - \lpar^\pm\vbloc}{2}\rt)\rt|\\
&= |\vzp^\pm(\vr_0+\vrp) - \vzp^\pm(\vr_0)|,
\label{eq:lpar}
\end{align}
where $\vbloc=\vBloc/|\vBloc|$ is the unit vector along the 
``local mean field'' $\vBloc\doteq\vB_0 + [\vbp(\vr_0)+\vbp(\vr_0+\vrp)]/2$. 
Note that $\lpar^\pm$ is a random quantity, {\em not} a parameter 
(unlike $\lambda$).

At each scale $\lambda$, the joint probability distribution fuction (PDF) 
$P(\dzp^+,\dzp^-,\theta,\lpar^+,\lpar^-|\lambda)$ contains all the information 
one customarily requires to characterize the structure of Alfv\'enic turbulence. 
As we only consider ``balanced'' turbulence, with equal mean injected power 
in the $+$ and $-$ fluctuations, $P$ is symmetric with respect 
to the $+$ and $-$ variables. We will use the $+$ mode wherever we need to make a choice.  
Imbalance leads to further interesting complications, left for future investigations. 

\begin{figure}
\vskipfig
\includegraphics[width=8cm]{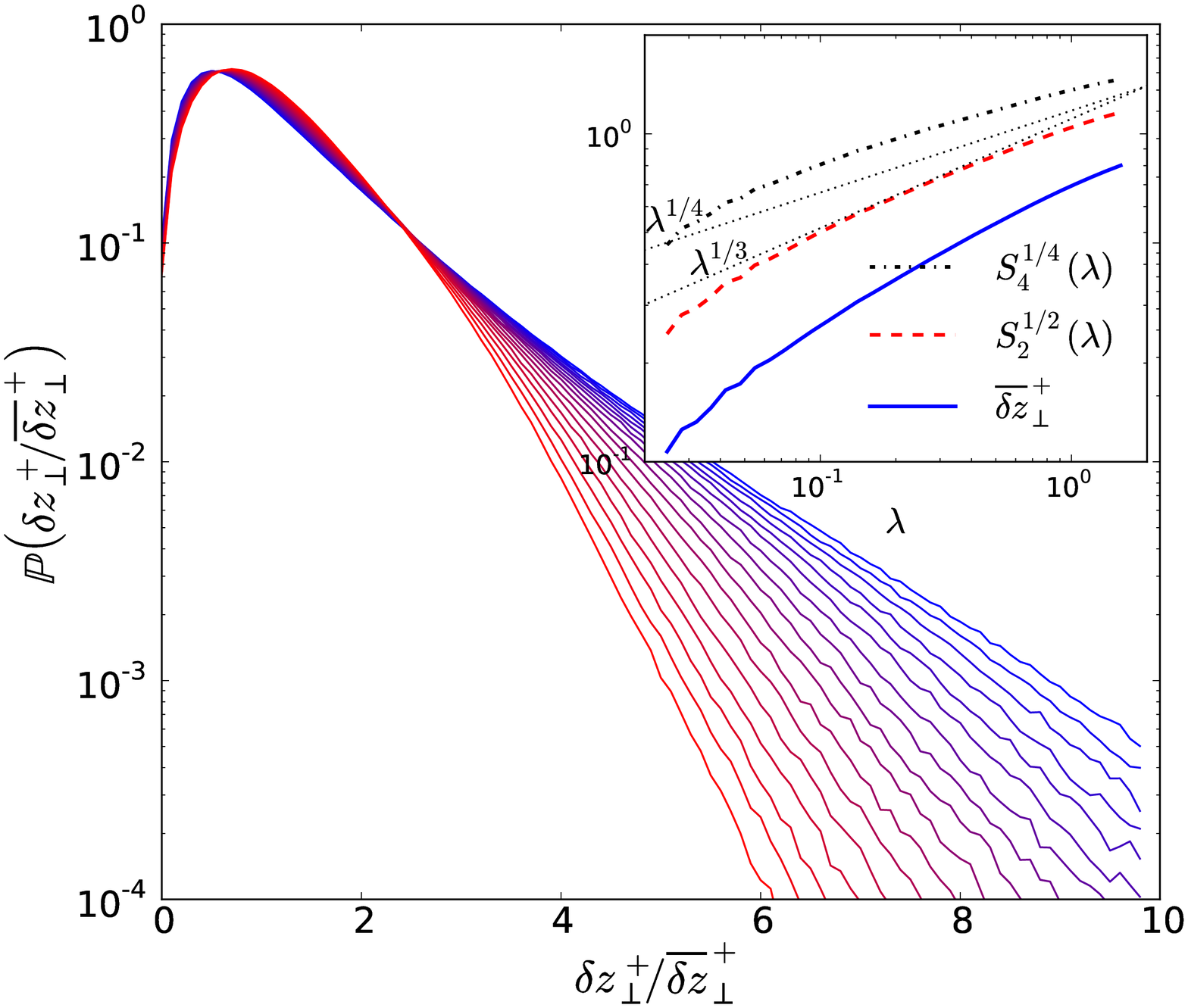}
\vskipfig
\caption{PDF of $\dzp^+$ rescaled to $\dzg^+\doteq\exp\la\ln\dzp^+ | \lambda\ra$, 
for scales from $\lambda=0.094$ (blue/dark) to $\lambda=0.92$ (red/light). 
{\em Inset:} the rms (2nd-order) increment 
$S_2^{1/2}(\lambda) \doteq \la(\dzp^+)^2|\lambda\ra^{1/2}$ (red dashed line), 
the 4th-order increment, 
$S_4^{1/4}(\lambda) \doteq \la(\dzp^+)^4|\lambda\ra^{1/4}$ (black dash-dotted line), 
and the ``typical'' increment $\dzg^+$ (blue solid line); 
the slopes $\lambda^{1/4}$ \citep{Boldyrev06} and
$\lambda^{1/3}$ \citep{GS95} 
are given for reference; all increments are normalized to the overall 
rms fluctuation level.}
\label{fig:dz}
\vskipfig
\end{figure}

\section{Numerical Experiment}

We solved \eqsref{eq:RMHD} using the code described in 
\citet{Chen11} in a triply periodic box of resolution $1024^3$. 
In the code units, $v_A = 1$ and the box length $=2\pi$ in each direction.  
The RMHD equations are invariant with respect to simultaneous rescaling 
$z \rightarrow a z$, $v_A \rightarrow a v_A$ for arbitrary $a$.
Therefore, although in code units the box is cubic and
$\dzp^\pm/v_A \sim 1$, in fact the box is much longer in the parallel 
than in the perpendicular direction and the fluctuation amplitudes are much 
smaller than $v_A$, while the linear and nonlinear terms remain comparable.
The energy was injected via white-noise forcing at $\kperp = 1,2$ and $\kpar = 1$ 
and dissipated by perpendicular hyperviscosity 
($\nu_\perp \nabla_\perp^8$ with $\nu_\perp = 2 \times 10^{-17}$) 
and Laplacian viscosity in $z$ ($\nu_z \dd^2 / \dd z^2$ with 
$\nu_z = 1.5\times 10^{-4}$; this is needed for numerical stability 
and has been checked to dissipate a negligible fraction of energy). 
The mean injected power was $\epsilon^\pm =1$ (balanced, strong turbulence). 
The forcing was purely in velocity; the magnetic field was not directly forced
(we have checked that when the two 
Elsasser fields are forced independently, all results reported below continue to hold).

The field increments \exref{eq:dz}, angles \exref{eq:theta} and parallel scales 
\exref{eq:lpar} were calculated for 32 logaritmically spaced scales, 
of which 17 were in the inertial range $0.094\le\lambda\le0.92$. 
For each $\lambda$, $10^6$ point separations were generated by choosing 
a random initial point $\vr_0$ on the grid and a random direction for $\vrp$ 
uniformly distributed in angle over a circle of radius $\lambda$ in the 
perpendicular plane. For each $\lambda$, the joint PDF $P$ was 
averaged over 10 such samples of $10^6$, 
from snapshots separated by approximately one large-scale eddy turnover~time.

\section{Results}

\subsection{Intermittency and Lack of Scale Invariance} 

A standard question of all turbulence studies is how the increments 
$\dzp^+$ depend on $\lambda$. 
As we anticipated above, the answer depends on which moment of the distribution 
$P(\dzp^+|\lambda)$ we choose to calculate. 
As shown in \figref{fig:dz} (inset) 
the rms increment $S_2^{1/2}(\lambda)\doteq\la(\dzp^+)^2|\lambda\ra^{1/2}$,  
based on the second-order structure function $S_2(\lambda)$,  
has a scaling between $\lambda^{1/3}$ 
(\citealt{GS95}'s $\kperp^{-5/3}$) and $\lambda^{1/4}$ 
(\citealt{Boldyrev06}'s $\kperp^{-3/2}$), with the usual difficulty 
of distinguishing between two very close exponents in a finite-resolution 
simulation. In contrast, 
the geometric, rather than aritmetic, mean $\dzg^+\doteq\exp\la\ln\dzp^+ | \lambda\ra$, 
perhaps better representing the ``typical realization,'' has a steeper scaling, 
whereas the ``fourth-order increment'' 
$S_4^{1/4}(\lambda)\doteq\la(\dzp^+)^4|\lambda\ra^{1/4}$ has a 
shallower one. The distribution is clearly not 
scale-invariant, as is made manifest by \figref{fig:dz}, 
where we show $P(\dzp^+|\lambda)$ rescaled to $\dzg^+$ 
at each $\lambda$. The salient feature of this PDF (which may be consistent with 
a lognormal, \citealt{Zhdankin12}, or a log-Poisson, \citealt{Chandran14}, distribution) 
is that it broadens at smaller $\lambda$---a classic case 
of intermittency understood as scale dependence of the distribution's shape.  

Other interesting quantities: $\theta$, $\lpar^\pm$, $\tA^\pm$, $\tnl^\pm$, etc.,  
also have intermittent, non-scale-invariant distributions. Let us focus 
on the two characteristic times. 

\begin{figure*}
\vskipfig
\includegraphics[width=8cm]{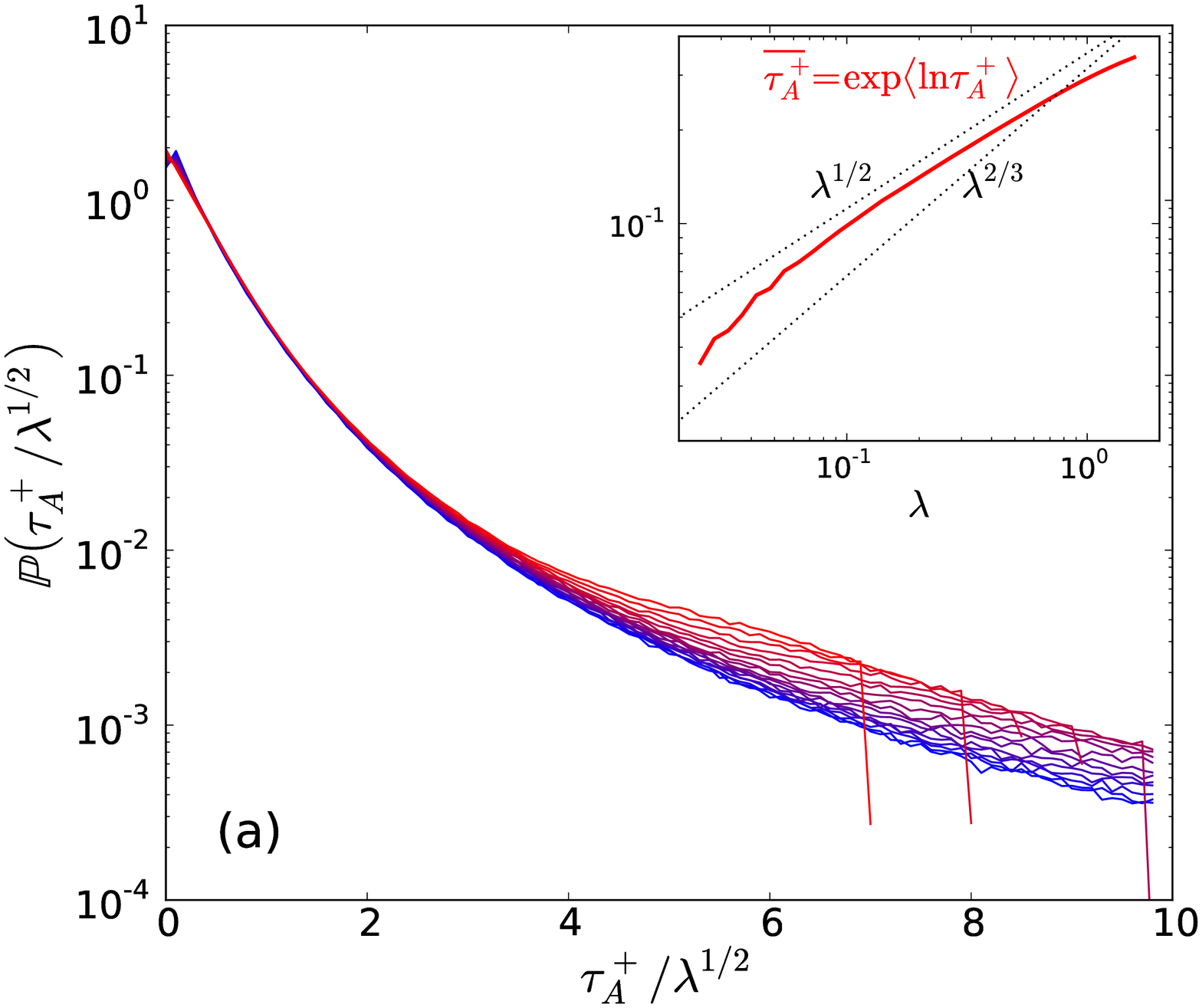}
\includegraphics[width=8cm]{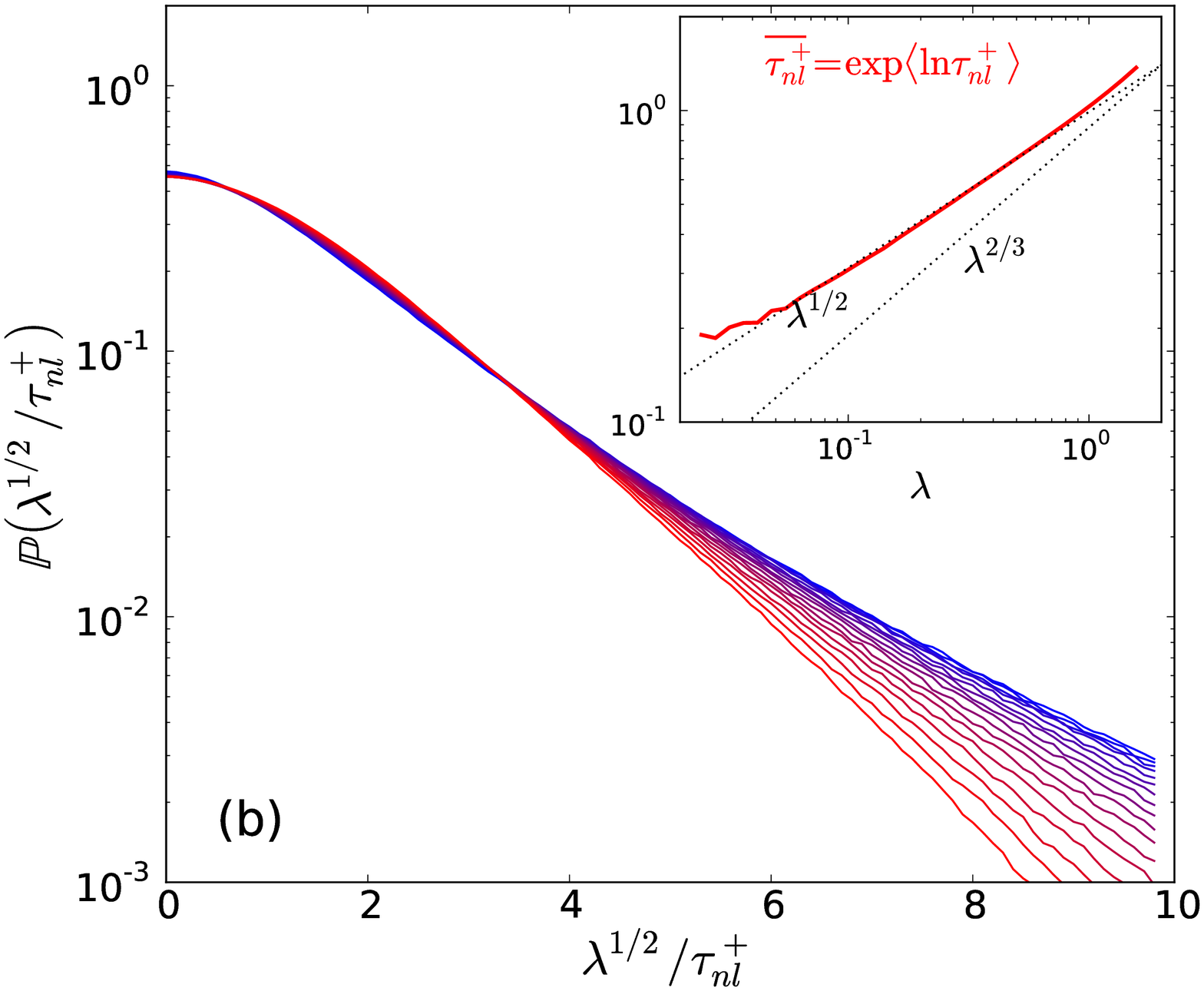}
\vskipfig
\caption{PDFs of (a) $\tA^+\doteq\lpar^+/v_A$ and (b) $(\tnl^+)^{-1}$ [\eqref{eq:tnl}], 
rescaled by $\lambda^{1/2}$, 
for scales from $\lambda=0.095$ (blue/dark) to $\lambda=0.92$ (red/large).
{\em Insets:} ``Typical times'' (a)  $\tAg\doteq\exp\la\ln\tA^+\ra$ and 
(b) $\tnlg\doteq\exp\la\ln\tnl^+\ra$ vs.\ $\lambda$; 
$\lambda^{1/2}$ and $\lambda^{2/3}$ scalings are shown for~reference.
\label{fig:tau}}
\vskipfig
\end{figure*}

\subsection{Alfv\'en Time and Nonlinear Time} 

The distribution of $\tA^\pm = \lpar^\pm/v_A$ is 
simply the distribution of the parallel coherence length. Its geometric mean 
is shown in \figref{fig:tau}(a, inset) and appears consistent with the scaling 
$\tAg^+\doteq\exp\la\ln\tA^+\ra \propto \lambda^{1/2}$, 
which is the relationship between the parallel and perpendicular 
scales that would follow from Boldyrev's phenomenology 
($\dzp^\pm\propto \lambda^{1/4} \propto (\lpar^+)^{1/2}$, \citealt{Boldyrev06}).\footnote{A more 
traditional way of extracting parallel scalings (corresponding 
to what is in fact done in the solar wind, \citealt{Horbury08,Podesta09,Wicks10,Chen11}) 
is to define {\em parallel} increments
$\tdzp^\pm \doteq |\vzp^\pm(\vr_0 + \lpar\vbloc) - \vzp^\pm(\vr_0)|$,
where $\vbloc$ is the local field direction at $\vr_0$ and $\lpar$ is a parameter, 
not a random variable. The rms of these increments is
$\la(\tdzp^+)^2 | \lpar\ra^{1/2} \propto \lpar^{1/2}$ \citep{Chen11}, 
which is reassuring as, replacing in \eqref{eq:eps} $\dzp^\pm\to\tdzp^\pm$, 
$\lpar^\pm\to\lpar$ and averaging, we get $\la(\tdzp^\pm)^2\ra\sim \lpar \la\varepsilon\ra/v_A$, 
where the mean injected power $\la\varepsilon\ra$ is certainly independent of scale.}
We see that it holds without being weighted 
by the fluctuation amplitude, i.e., it is a measure of the 
prevailing spatial anisotropy in the system. 
The PDFs of the rescaled quantity $\tA^+/\lambda^{1/2}$ for a range of $\lambda$ are 
shown in \figref{fig:tau}(a): at smaller $\tA^+/\lambda^{1/2}$ 
(i.e., relatively shorter $\lpar^+$), 
there appears to be a scale-invariant collapse, but at larger values, the PDF becomes 
non-scale-invariant---with a systematically shallower tail at larger $\lambda$.


The geometric mean of the nonlinear time is shown in
\figref{fig:tau}(b, inset) and, like $\tAg^+$, scales as 
$\tnlg^+ \doteq \exp\la\ln\tnl^+ | \lambda\ra \propto \lambda^{1/2}$. 
Note that the presence of the alignment angle $\theta$ in the definition \exref{eq:tnl} 
of $\tnl^\pm$ is essential because it reduces the strength of the nonlinear interaction 
in a scale-dependent way. 
The PDFs of the rescaled inverse nonlinear time, $\lambda^{1/2}/\tnl^+$, are shown 
in \figref{fig:tau}(b). There is approximate (but clearly not perfect) scale invariance 
at small values of the rescaled quantity (i.e., relatively longer $\tnl$), 
and a very non-scale-invariant tail at larger values, systematically shallower 
at smaller~$\lambda$. 

\subsection{Refined Critical Balance} 

\begin{figure}
\vskipfig
\includegraphics[width=8cm]{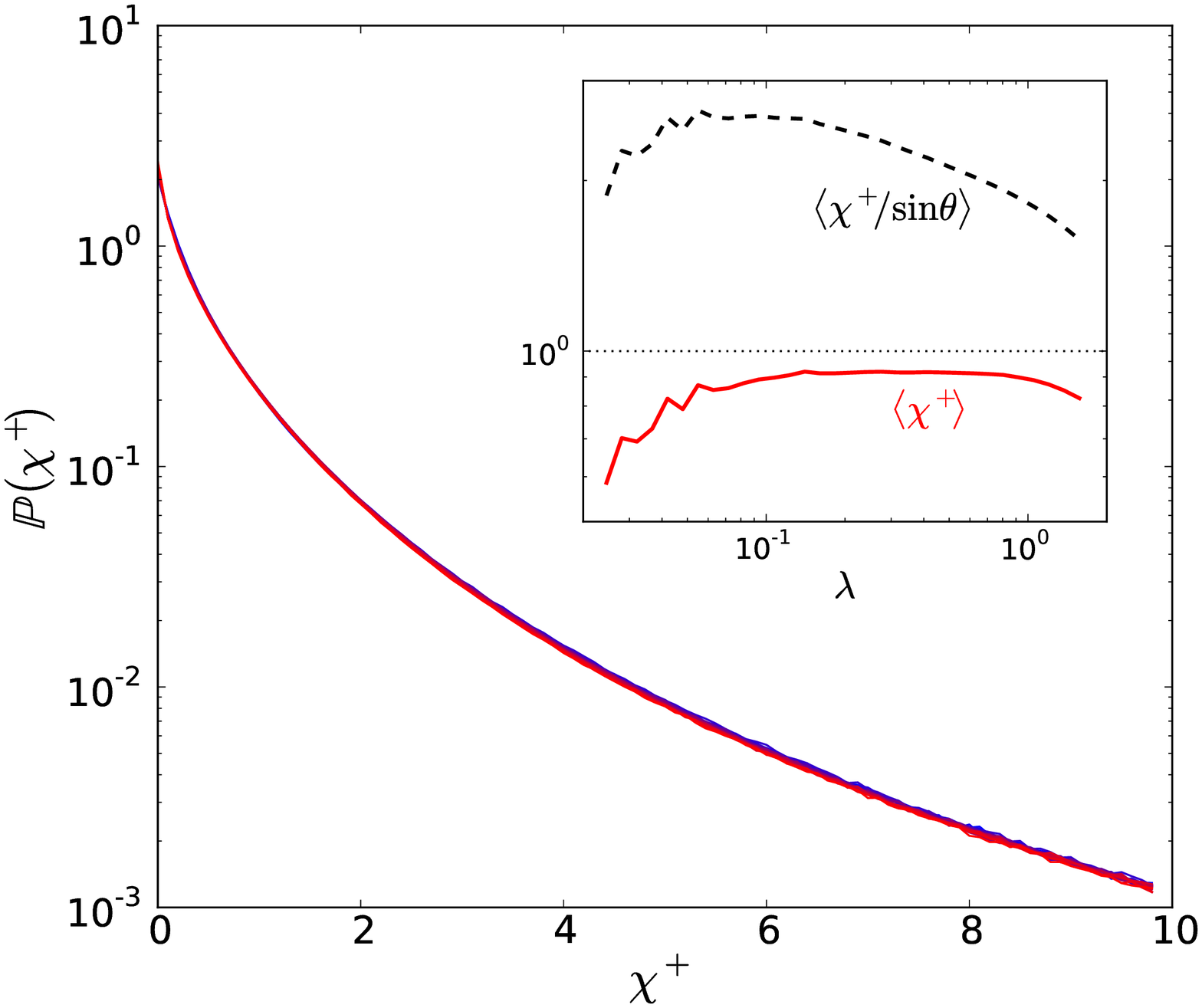}
\vskipfig
\caption{PDF of $\chi^+$ (defined by \eqref{eq:chi}) 
for scales from $\lambda=0.094$ (blue/dark) to $\lambda=0.92$ (red/large). 
Data collapse is nearly perfect. 
{\em Inset:} the mean nonlinearity parameter $\la\chi^+\ra$ vs.\ $\lambda$ (red/solid) 
and the same without account for alignment, $\la\chi^+/\sin\theta\ra$ (black/dashed).} 
\label{fig:chi}
\vskipfig
\end{figure}

The behaviour of the distribution of the nonlinear time
fits neatly with that of the distribution of the Alfv\'en time. 
The cores of both distributions 
(roughly, $\tA^+/\lambda^{1/2} \lesssim 3$ and $\lambda^{1/2}/\tnl^+ \lesssim 3$
in \figref{fig:tau}) are close to being scale invariant. 
On the other hand, their tails vary with $\lambda$ in opposite senses, 
with the tail of $\tA^\pm/\lambda^{1/2}$ ($\lambda^{1/2}/\tnl^\pm$) 
becoming steeper (shallower) as $\lambda$ decreases.  Because of this, the distribution 
of their product $\chi^\pm$, defined in \eqref{eq:chi}, 
does not change at all: $P(\chi^+|\lambda)$, shown in \figref{fig:chi}, 
is independent of $\lambda$ across the inertial range and all its moments 
are constant: e.g., $\la\chi^+ | \lambda\ra$ is shown in the inset of \figref{fig:chi}  
(alongside it, we show the mean nonlinearity parameter without 
the $\sin\theta$ factor, $\la\chi^+/\sin\theta | \lambda\ra$; it is not 
scale-independent, so the alignment is an essential ingredient of the RCB). 

That the nonlinearity parameter $\chi^\pm$ 
has a scale-invariant distribution is the main result of this Letter.  
This is due to the fundamental physical connection 
between the parallel and perpendicular structure of turbulent 
fluctuations---they cannot remain coherent beyond a parallel distance that information 
propagates at the Alfv\'en speed during one perpendicular nonlinear decorrelation time, 
$\tA^\pm \sim \tnl^\pm$. 

\subsection{Alignment}

\begin{figure}
\vskipfig
\includegraphics[width=8cm]{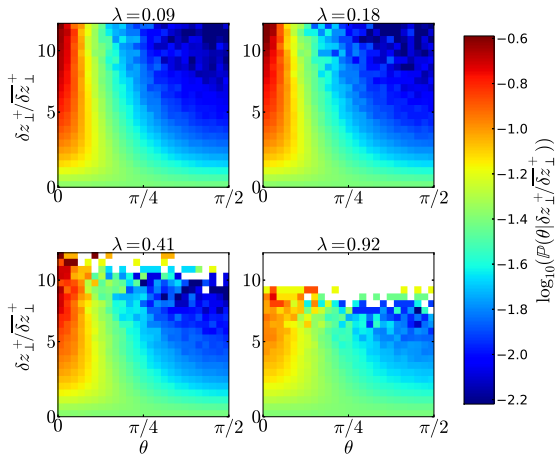}
\vskipfig
\caption{PDF of the alignment angle $\theta$ conditional on the 
fluctuation amplitude $\dzp^+$ relative to ``typical'' value 
$\dzg^+\doteq\exp\la\ln\dzp^+ | \lambda\ra$, 
viz., $P(\sin\theta | \dzp^+/\dzg^+, \lambda)$, 
plotted for four representative scales $\lambda$ (as shown).}
\label{fig:alignment}
\vskipfig
\end{figure}

The role of alignment in giving rise to the RCB deserves further discussion. 
At every scale $\lambda$, the fluctuation amplitude $\dzp^\mp$ and the alignment 
angle $\theta$ turn out to be {\em anticorrelated} (cf.~\citealt{Beresnyak06}). 
This is best demonstrated by the 
conditional PDF $P(\sin\theta | \dzp^+/\dzg^+, \lambda)$, shown in \figref{fig:alignment}. 
We see that fluctuations whose amplitudes are large relative to the ``typical'' 
value $\dzg^+$ (i.e., those giving rise to the shallow intermittent tails 
manifest in \figref{fig:dz}) tend to be well aligned, 
whereas the weaker fluctuations ($\dzp^+/\dzg^+\lesssim 1$) 
are unaligned. The alignment of the stronger fluctuations appears to get 
statistically ``tighter'' at smaller scales. 

Thus, for the stronger fluctuations, the nonlinear interaction 
is reduced by alignment more than for the weaker ones.  
We find the approximately scale-invariant core of the distribution 
of $\lambda^{1/2}/\tnl^+$ in \figref{fig:tau}(b) 
to contain simultaneously smaller $\theta$ but relatively larger $\dzp^-$, 
so it is the more aligned fluctuations that 
give rise to the \citet{Boldyrev06} scaling $\dzp^+\propto\lambda^{1/4}$, 
as expected. Note, however, that  
the anticorrelation between alignment and amplitude is somewhat at odds with 
Boldyrev's intuitive interpretation of the alignment angle as determined by the maximal angular 
wander within any given fluctuation ($\theta\sim\dbp/B_0$), but rather 
suggests that alignment might be caused by dynamical shearing of a weaker 
Elsasser field by a stronger one (\citealt{Chandran14}; the anticorrelation 
holds for both the weaker and the stronger of the 
two Elsasser fields, but is slightly more pronounced if \figref{fig:alignment} is replotted 
for $P(\sin\theta | \dzp^{\rm (max)}/\dzg^{\rm (max)}, \lambda)$
with $\dzp^{\rm (max)}$ the locally stronger field). 
Qualitatively, this is why measures of alignment weigted by the energy (or higher powers 
of fluctuation amplitudes) exhibit stronger scale dependence \citep{Beresnyak09,Mallet14}. 

All of these statements must be accompanied by the acknowledgment that 
a debate continues as to whether the tendency to alignment 
in Alfv\'enic turbulence survives at asymptotically small scales, 
with numerical simulations at resolutions up to $4096^3$ 
falling short of an indisputable outcome \citep{Perez12,Beresnyak14}. 
What does, however, appear to be solidly the case is that Alfv\'enic fluctuations 
over at least the first two decades below the outer scale do exhibit 
alignment, even if transiently (cf.\ \citealt{Podesta09jgr,Chen12,Wicks13prl,Wicks13apj}), 
that they do this in a systematic, scale- and 
amplitude-dependent fashion and, as argued above, 
that this effect must be taken into 
account in interpreting what it means, statistically, for these 
fluctuations to be in a critically balanced state. 
The possible change of regime at even smaller 
scales \citep{Beresnyak14} is left outside the scope of the present work. 

\section{Conclusion} 

The results presented above imply that the 
structure of Alfv\'enic turbulence is set by two fundamental effects: 
the critical balance, {\em which occurs in a scale-invariant fashion} 
(probably due to the upper limit on the parallel coherence length of turbulent 
fluctuations imposed by causality over a nonlinear decorrelation time), 
and systematic alignment of the higher-amplitude fluctuations 
(probably due to dynamical mutual shearing of Elsasser fields). 
The first of these results suggests that critical balance---quantitatively 
amounting, as we have argued, to the RCB conjecture---is the 
most robust and reliable of the physical principles undepinning theories
of Alfv\'enic turbulence. 

While scale-dependent alignment of inertial-range fluctuations in the solar wind 
is still in question \citep{Podesta09jgr,Chen12,Wicks13prl,Wicks13apj},
measurements of the anisotropy/alignment/intermittency of these fluctuations 
directed at the verification of the RCB might help establish whether numerical 
and real plasma turbulence share the key 
structural properties and whether, therefore, debates and insights arising 
from the former have a useful contribution to make to the understanding 
of the latter.

 
We are grateful to A.~Beresnyak, C.~H.~K.~Chen, S.~C.~Cowley, T.~S.~Horbury, 
J.~C.~Perez, and R.~T.~Wicks for many useful discussions of MH turbulence.
AM was supported in part by STFC (UK), 
BDGC by NASA grants NNX11AJ37G and NNX12AB27G, NSF grant
AGS-1258998. 
Simulations reported here used XSEDE, which is supported by 
the US NSF Grant ACI-1053575.

\bibliographystyle{mn2e}
\bibliography{msc_MNRAS}

\label{lastpage}
\end{document}